\documentclass[aps,twocolumn,prl,superscriptaddress,amsmath,amssymb]{revtex4-2} 
\usepackage{dcolumn}
\usepackage{bm}
\usepackage{amsmath}
\usepackage{txfonts}
\usepackage[T1]{fontenc}
\usepackage{xspace}
\usepackage{ulem}
\usepackage{comment}
\newcommand{\bras}[1]{\langle#1\rvert}
\newcommand{\kets}[1]{\lvert#1\rangle}

\newcommand{\bra}[1]{\left<#1\right|}
\newcommand{\ket}[1]{\left|#1\right>}

\newcommand{\means}[1]{\langle#1\rangle}

\newcommand{\meanss}[1]{\langle\!\langle#1\rangle\!\rangle}

\ifx\pdfoutput\undefined
\usepackage[dvipdfmx]{graphicx}
\usepackage[dvipdfmx]{hyperref}
\usepackage[dvipdfmx]{color}
\usepackage[dvipdfmx]{xcolor}
\else
\usepackage{graphicx}
\usepackage{hyperref}
\usepackage{color}
\usepackage{xcolor}
\fi
\hypersetup{colorlinks=true,linkcolor=blue,citecolor=blue,urlcolor=blue}
\begin{document}
\let\emph\textit

\title{
  Field-driven spatiotemporal manipulation of Majorana zero modes in Kitaev spin liquid
}
\author{Chihiro Harada}
\affiliation{
  Department of Physics, Tohoku University, Sendai 980-8578, Japan
}
\author{Atsushi Ono}
\affiliation{
  Department of Physics, Tohoku University, Sendai 980-8578, Japan
}
\author{Joji Nasu}
\affiliation{
  Department of Physics, Tohoku University, Sendai 980-8578, Japan
}

\date{\today}
\begin{abstract}
  The Kitaev quantum spin liquid possesses two fractional quasiparticles, itinerant Majorana fermions and localized visons.
  It provides a promising platform for realizing a Majorana zero mode trapped by a vison excitation.
  This local mode behaves as a non-Abelian anyon capable of applications to quantum computation.
  However, creating, observing, and manipulating visons remain challenging even in the pristine Kitaev model.
  Here, we propose a theory to control visons enabled by a time-dependent local magnetic field in the Kitaev spin liquid.
  Examining the time evolution of the magnetic state, we demonstrate that a vison follows a locally applied field sweeping in the system.
  We clarify that one can move a vison accompanied by a Majorana zero mode by choosing the velocity and shape of the local field appropriately.
  In particular, the controllability of visons using local fields shows nonlinear behavior for its strength, which originates from interactions between Majorana fermions and visons.
  The present results suggest that itinerant Majorana fermions other than zero modes play a crucial role in vison transport.
  Our finding will offer a guideline for controlling Majorana zero modes in the Kitaev quantum spin liquid.
\end{abstract}
\maketitle


Electron correlations in solids bring about exotic ground states with strong quantum entanglement, and resultant elementary excitations possess completely different natures from electrons.
Amongst others, elementary excitations behaving like independent particles emergent from an electron divided are called fractional quasiparticles.
Elucidating their properties is one of the main subjects in modern condensed matter physics. 
A representative example of fractional quasiparticles is a Majorana fermion, which is identical to its own antiparticle~\cite{Majorana1937}.
The fractionalization into multiple Majorana quasiparticles can cause non-Abelian anyonic statistics between them, which is applied to topological quantum computation~\cite{kitaev2003fault,freedman2003topological,Nayak2008,Ahlbrecht2009quantum_computer}.
Therefore, manifestations of Majorana quasiparticles have been intensively investigated in many-body quantum states, such as unconventional superconductors~\cite{Sarma2006,Sato2009topological,Fu2008Superconducting,Sato2016Majorana,Sato_rev2017} and fractional quantum Hall systems~\cite{moore1991nonabelions,Stormer1999,Read2000,banerjee2018observation}.

A quantum spin model proposed by A.~Kitaev offers another platform of Majorana fermions~\cite{Kitaev2006}.
This model, composed of simple interactions between $S=1/2$ quantum spins, exhibits a quantum spin liquid (QSL) as its ground state.
Elementary excitations from the QSL are described by two fractional quasiparticles: Majorana fermions and local excitations termed visons~\cite{RevModPhys.87.1,Hermanns2018rev,Knolle2019rev,takagi2019rev,Janssen_2019rev,Motome2020rev,Trebst2022rev}.
Under magnetic fields, the Majorana fermion system is topologically nontrivial, which results in a Majorana fermion with zero energy, called a Majorana zero mode, bound by each vison.
This composite quasiparticle behaves as a non-Abelian anyon similar to a zero-energy Majorana state trapped by a vortex in superconductors.
Recent extensive investigations have clarified that the Kitaev model can be realized in transition metal compounds~\cite{PhysRevLett.102.017205,PhysRevLett.105.027204,PhysRevLett.110.097204,PhysRevLett.113.107201,PhysRevLett.112.077204,Winter2016}, metal-organic frameworks~\cite{Yamada2017}, and cold atom systems~\cite{Duan2003,Manmana2013,gorshkov2013kitaev,Fukui2022}. 
In particular, the half-integer quantized thermal Hall effect inherent to topological Majorana fermions has been reported to be observed in the ruthenium compound $\alpha$-RuCl$_3$ under magnetic fields~\cite{kasahara2018majorana,yokoi2021half,Yamashita2020_Sample_dependence,bruin2022robustness}.
While the presence of the quantization has still been under debate~\cite{Hentrich2019,czajka2022planar,Lefrancois2022}, this material is believed to be a promising candidate for the Kitaev QSL with non-Abelian anyons.
For applications of the Kitaev systems to quantum computing, one needs to create, observe and manipulate vison excitations as desired.
Several proposals for observing visons have been made via scanning tunneling microscope (STM) and atomic force microscope (AFM) measurements~\cite{Feldmeier2020,Konig2020,Pereira2020,Chen2020Impurity,Carrega2020,Udagawa2021,Jang2021,Bauer2023} and interferometry for Majorana edge modes~\cite{Klocke2021,Wei2021,Liu2022AnyonGeneration} in this context.
However, less is known about how to manipulate a vison with a Majorana zero mode, even in the pristine Kitaev model.

In this Letter, we theoretically propose a way to control visons using a local magnetic field in the Kitaev QSL.
We introduce the Kitaev model under a weak uniform magnetic field.
This field makes each vison excitation accompanied by a Majorana fermion but does not invest itinerant nature to visons.
Here we show that visons can be spatiotemporally manipulated by applying a local field sweeping across the system.
We also reveal optimal conditions of the moving velocity and intensity of the local field to control a vison.
Furthermore, we find a nonlinearity for the local field caused by interactions between itinerant Majorana fermions and visons, which are crucial for vison manipulations.


\begin{figure}[t]
  \begin{center}
  \includegraphics[width=\columnwidth,clip]{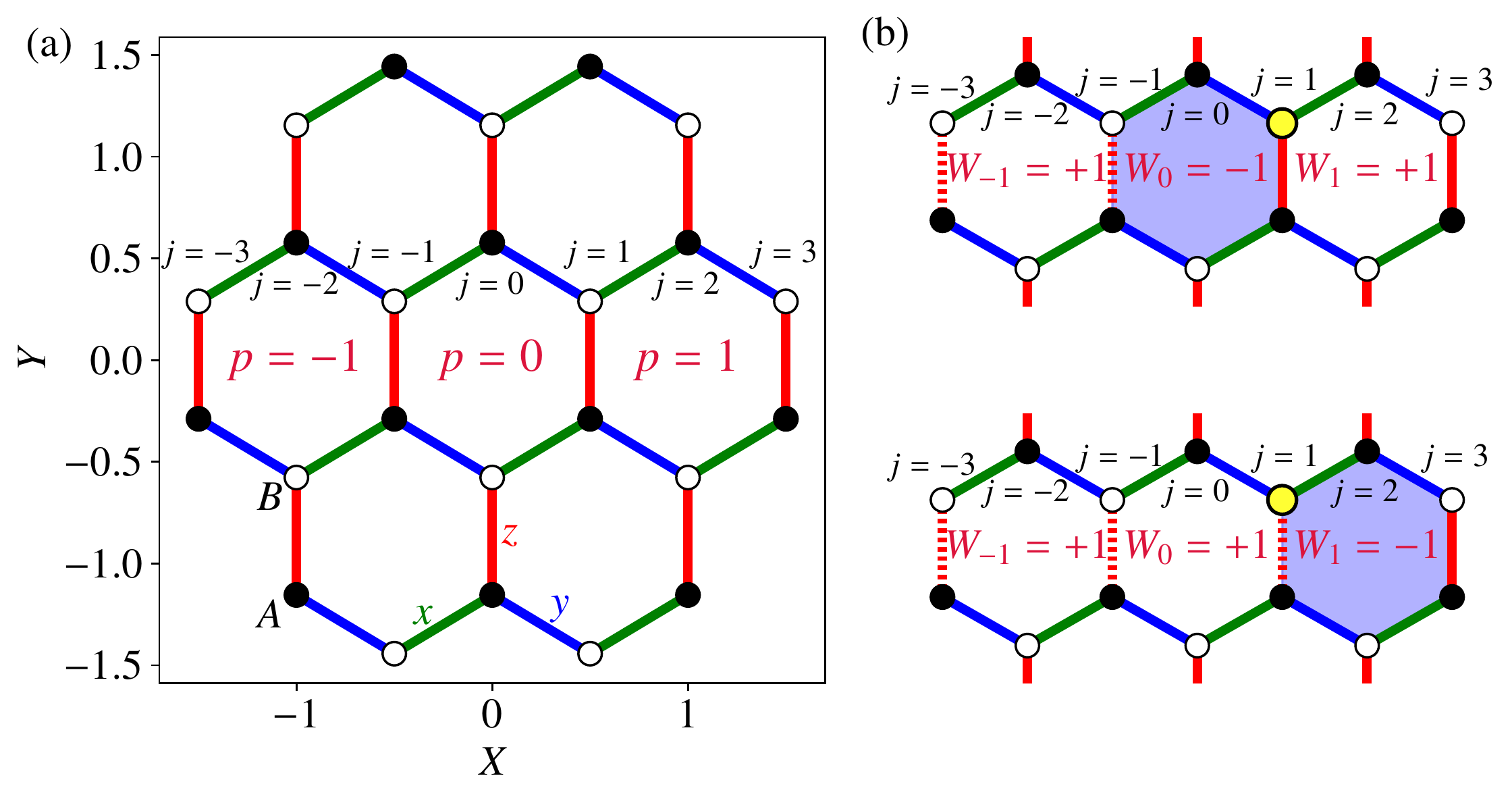}
  \caption{
    (a) Honeycomb lattice cluster with open boundaries including $N=24$ spins on which the Kitaev model is defined.
    Blue, green, and red lines represent the $x$, $y$, and $z$ bonds, respectively, and filled (open) circles stand for sites belonging to the $A$ ($B$) sublattice.
    The origin of the spatial coordinate is taken at the center of the cluster, and each hexagon is labeled by $p$ with the central hexagon as $p=0$ and numbered along the $X$ direction.
    In a similar manner, sites are numbered with $j$ along $xy$ bonds toward the $X$ direction such that the uppermost site of the central hexagon with $p=0$ is $j=0$.
    (b) Two flux configurations with a vison at $p=0$ and $1$ associated with $Z_2$ variables with $\eta_r=-1$ on the $z$ bonds shown by dashed lines.
    The matrix element of $S^z$ on the yellow site ($j=1$) is nonzero between two flux configurations.
  }
  \label{fig:lattice}
  \end{center}
\end{figure}

We start from the Kitaev model under a uniform magnetic field as follows~\cite{Kitaev2006}:
\begin{align}
  {\cal H}_{K}=-J\sum_{\means{jj'}_{\gamma}}S_j^\gamma S_{j'}^\gamma
  -\kappa \sum_{\meanss{jj'j''}_{\gamma\gamma'\gamma''}}S_j^{\gamma}S_{j'}^{\gamma'}S_{j''}^{\gamma''},
  \label{eq:HamilS}
\end{align}
where $S_j^\gamma$ is the $\gamma$ component of an $S=1/2$ spin at site $j$ on a honeycomb lattice.
The first term represents the bond-dependent Ising-type interaction between spins on nearest neighbor (NN) $\gamma$ bond $\means{jj'}_{\gamma}$ ($\gamma=x,y,z$) [Fig.~\ref{fig:lattice}(a)] with the exchange constant $J$, and the second term stands for 
an effective field leading to a chiral spin liquid with a tolopogically nontrivial gap.
The latter is given as the product of three spins on neighboring three sites $\meanss{jj'j''}_{\gamma\gamma'\gamma''}$ consisting of $\gamma$ and $\gamma''$ bonds with $\gamma'$ being neither $\gamma$ nor $\gamma''$.
This term has been proposed to originate from higher-order perturbations for the Zeeman term~\cite{Kitaev2006} and $\Gamma'$ interactions~\cite{Takikawa2019,Takikawa2020} and has been discussed as an independent term from magnetic fields ~\cite{Takikawa2022,Chen2023,Takahashi2022pre,Kao2023pre}.
By performing the Jordan-Wigner transformation along chains composed of $x$ and $y$ bonds, we can rewrite Eq.~\eqref{eq:HamilS} using two Majorana fermions $c_j$ and $\bar{c}_j$ at each site as~\cite{PhysRevB.76.193101,PhysRevLett.98.087204,1751-8121-41-7-075001,PhysRevLett.113.197205,PhysRevB.92.115122,Minakawa2020}
\begin{align}
  &{\cal H}_{K}=-\frac{J}{4}\sum_{[jj']_x}ic_j c_{j'}-\frac{J}{4}\sum_{[jj']_y}ic_j c_{j'}
  -\frac{J}{4}\sum_{[jj']_z}\eta_r ic_j c_{j'}\nonumber\\
  &-\frac{\kappa}{8}\sum_{\meanss{jj'j''}_{zxy}}\eta_r ic_j c_{j''}
  -\frac{\kappa}{8}\sum_{\meanss{jj'j''}_{xyz}}\eta_r ic_j c_{j''}
  -\frac{\kappa}{8}\sum_{\meanss{jj'j''}_{yzx}}ic_j c_{j''},
  \label{eq:Hamilc}
\end{align}
where the spin operator on sublattice $A$ ($B$) is given by $(S_j^x,S_j^y,S_j^z)=\frac{1}{2}(c_j\tau_j,-\bar{c}_j\tau_j,i c_j\bar{c}_j)$ [$\frac{1}{2}(\bar{c}_j\tau_j,-c_j\tau_j,i \bar{c}_j c_j)$] with $\tau_j=\prod_{j'<j}\left(-2S_{j'}^z\right)$ [see Fig.~\ref{fig:lattice}(a)].
Here, $[jj']_\gamma$ is the ordered NN pair and $\eta_r =i\bar{c}_j \bar{c}_{j'}$ on $z$ bond $r$, where $j$ ($j'$) belongs to sublattice $A$ ($B$).
The quantity $\eta_r$ exists also on the $z$ bond of neighboring three sites $\meanss{jj'j''}_{zxy}$ and $\meanss{jj'j''}_{xyz}$.
Since $\eta_r$ commutes with ${\cal H}_{K}$ and satisfies $\eta_r^2=1$, it is a local $Z_2$ conserved quantity.
We also introduce a physical local conserved quantity as $W_p=2^6 \prod_{j\in p} S_j^{\gamma_j}$ on each hexagon plaquette $p$ with $\gamma_j$ being the bond type not belonging to the hexagon loop at site $j$.
This quantity relates to $\eta_r$ as $W_p=\eta_{r_1}\eta_{r_2}$, where $r_1$ and $r_2$ are the two $z$ bonds on the hexagon $p$.
The ground state of ${\cal H}_{K}$ belongs to the sector with all $W_p$ being $+1$, called flux-free. 
Then, the elementary excitations are described by two quasiparticles: itinerant Majorana fermions $c_j$ and local excitations flipping $W_p$ to $-1$.
The latter quasiparticles are termed visons.

Figure~\ref{fig:lattice}(b) shows two single-vison excitations, where a vison is located at the endpoint of the string crossing $z$ bonds with $\eta_r=-1$.
To manipulate visons, we need to consider an external field with nonzero matrix elements between the two configurations.
One of the simplest ways is introducing a local magnetic field for $S_1^z$ at the yellow site $j=1$ shown in Fig.~\ref{fig:lattice}(b)~\cite{Joy2022,Chen2023}.
Since $S_1^z$ anticommutes with $\eta_{r_0}$ being the $z$ bond connected with the site $j=1$, it causes the hopping of a vison between neighboring hexagon plaquettes.
We introduce the time-dependent local field ${\cal H}_h(t)=-\sum_j h_j(t)S_j^z$ for spatial and temporal control of visons.

In the present study, we apply the Hartree-Fock approximation to the Majorana fermion representation of ${\cal H}={\cal H}_K+{\cal H}_h(t)$; the third, fourth, and fifth terms in Eq.~\eqref{eq:Hamilc} are decoupled as 
$
i\bar{c}_{i}\bar{c}_{j}ic_{i}c_{j}\simeq
\means{i\bar{c}_{i}\bar{c}_{j}}ic_{i}c_{j}+i\bar{c}_{i}\bar{c}_{j}\means{ic_{i}c_{j}} -\means{ic_{i}\bar{c}_{j}}ic_{j}\bar{c}_{j}
-ic_{i}\bar{c}_{j}\means{ic_{j}\bar{c}_{j}}+\means{ic_{i}\bar{c}_{j}}ic_{j}\bar{c}_{i}+ic_{i}\bar{c}_{j}\means{ic_{j}\bar{c}_{i}}
$ with a constant.
The obtained mean-field Hamiltonian ${\cal H}_{\rm MF}(t)$ is given by a following bilinear form of Majorana fermions $\{\gamma_l\}=\{c_{1},c_{2},\cdots,c_{N},\bar{c}_{1},\bar{c}_{2},\cdots,\bar{c}_{N} \}$:
$\mathcal{H}_{\rm MF} = \frac{i}{4}\sum_{ll'}\gamma_{l}{\cal A}_{ll'}\gamma_{l'}+C$,
where $N$ is the number of sites, $C$ is a constant, and ${\cal A}$ is a $2N\times2N$ real skew symmetric matrix~\cite{suppl}.

We calculate the time evolution based on the von Neumann equation and solve it by the fourth-order Runge-Kutta method with the step size $\Delta t$~\cite{volkov1973collisionless,Tsuji_theoryof2015,Murakami_collective2020,Nasu2022}.
As an initial state at $t_{\rm in}=0$, we impose no local magnetic field with ${\cal H}_h(t_{\rm in})=0$, where $W_p$ and $\eta_r$ are conserved quantities.
Numerical calculations are performed with $\Delta t/J^{-1}=0.01$ ($\hbar=1$ is assumed) on the hexagon-type cluster with $N=1734$, 
which is straightforwardly obtained by extending the cluster shown in Fig.~\ref{fig:lattice}(a)~\cite{suppl}.
The width of the cluster for the $X$ direction is 33, where the length of the primitive translation vectors is unity.
We compute the expectation value of $W_p$ and local density of states (LDOS) $g_j(\varepsilon;t)$ for the Majorana fermions $c_j$ at site $j$, which is defined to satisfy the normalization $\int_0^\infty g_j(\varepsilon;t)d\varepsilon=1$.
The details are given in Supplemental Material~\cite{suppl}.


\begin{figure*}[t]
  \begin{center}
  \includegraphics[width=2\columnwidth,clip]{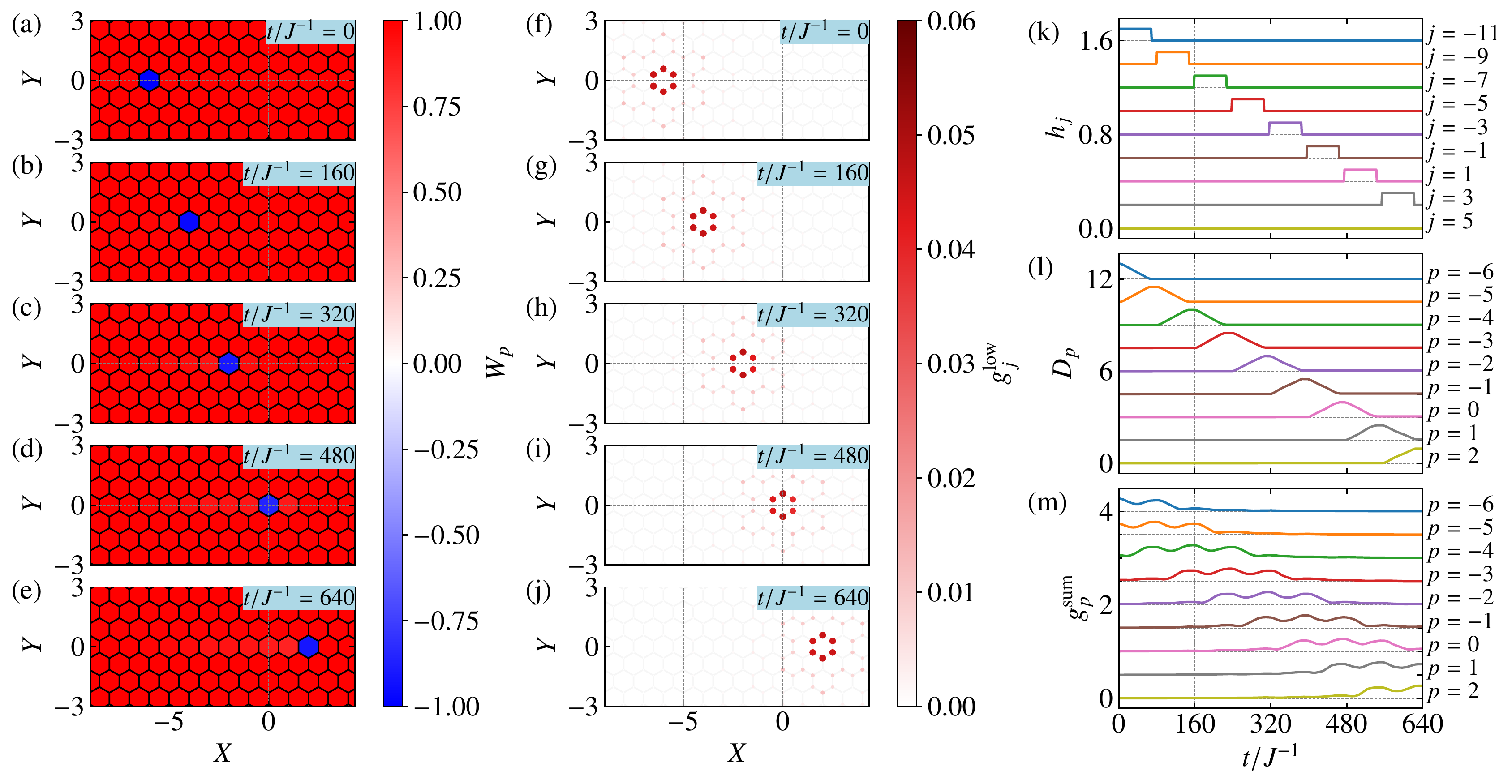}
  \caption{
    (a)--(e) Spatial map of $W_p$ under the time-dependent local field at (a) $t=0$, (b) $t/J^{-1}=160$, (c) $t/J^{-1}=320$, (d) $t/J^{-1}=480$, (e) $t/J^{-1}=640$.
    (f)--(j) Corresponding figures for low-energy LDOS $g_j^{\rm low}$.
    (j)--(m) Time evolutions of (j) the local fields $h_j$, (k) vison density $D_p$, and (m) low-energy LDOS $g_p^{\rm sum}$ on the hexagon plaquette $p$.
  }
  \label{fig:XYmap}
  \end{center}
\end{figure*}

Here, we show the results for the time evolution of a vison in the presence of spatially and temporally dependent local field ${\cal H}_h(t)$.
In the initial state, a vison is located at $(X,Y)=(p_{\rm in},0)$ with $p_{\rm in}=-6$, namely $W_{p_{\rm in}}=-1$ and other $W_p$ are $+1$ [see Fig.~\ref{fig:XYmap}(a)].
The local field is applied for the sites $j=2p_{\rm in}+1,2p_{\rm in}-1,\cdots, 5$, where $j_{\rm min}=2p_{\rm in}+1$ is the upper right site of the hexagon $p_{\rm in}$, as a shifted rectangular field $h_j(t)=A \mathcal{R}(t;(T+\Delta T)(j-j_{\rm min})/2,T)$ with the amplitude $A$ and delay time $\Delta T$ [see Fig.~\ref{fig:XYmap}(k)].
$\mathcal{R}(t;t_0,T)=\theta(t-t_0)\theta(t_0+T-t)$ is a rectangular function starting from $t_0$ with the width $T$ and $\theta(t)$ is the Heaviside step function.
We set the parameters as $A/J=0.1$, $T/J^{-1}=69$, $\Delta T/J^{-1}=10$, and $\kappa/J=0.05$.
Figures~\ref{fig:XYmap}(a)--\ref{fig:XYmap}(e) show the snapshots of the spatial map for $\means{W_p}$.
We find that the vison moves to the right side by following the motion of the local field and its shape is largely intact by time evolution.
Figure~\ref{fig:XYmap}(l) shows the time evolution of the vison density $D_p$ on the hexagons whose centers are located on the $Y=0$ line, where $D_p=(1-\means{W_p})/2$.
The NN hopping of a vison occurs while the rectangular local field is applied, and the vison almost perfectly moves to the neighboring site.

Figures~\ref{fig:XYmap}(f)--\ref{fig:XYmap}(j) show the snapshots of the low-energy LDOS $g_j^{\rm low}(t)$, where $g_j^{\rm low}(t)=\int_0^{\varepsilon_m} g_j(\varepsilon;t)d\varepsilon$ with $\varepsilon_m/J=10^{-3}$.
At $t=0$, the large spectral weight of the low-energy LDOS is observed at the corners of the hexagon with a vison, corresponding to a Majorana zero mode.
The zero mode clearly follows the vison moving to the right side while maintaining its spatial shape in time evolution.
This feature is also seen in Fig.~\ref{fig:XYmap}(m), showing the low-energy LDOS on the hexagon plaquette $p$, $g_p^{\rm sum}(t)=\sum_{j\in p}g_j^{\rm low}(t)$.
The present results suggest that a vison is accompanied by a Majorana zero mode even while a time-dependent local field drives a vison.
Note that this field-driven vison manipulation is sensitive to the parameters of the local field.
When one changes the value of $A$, the wavepacket of a vison spreads out and the quasiparticle picture collapses regardless of its propagation~\cite{suppl}.
The results imply that the shape of the local magnetic field is crucial in manipulating a vison.

\begin{figure}[t]
  \begin{center}
  \includegraphics[width=\columnwidth,clip]{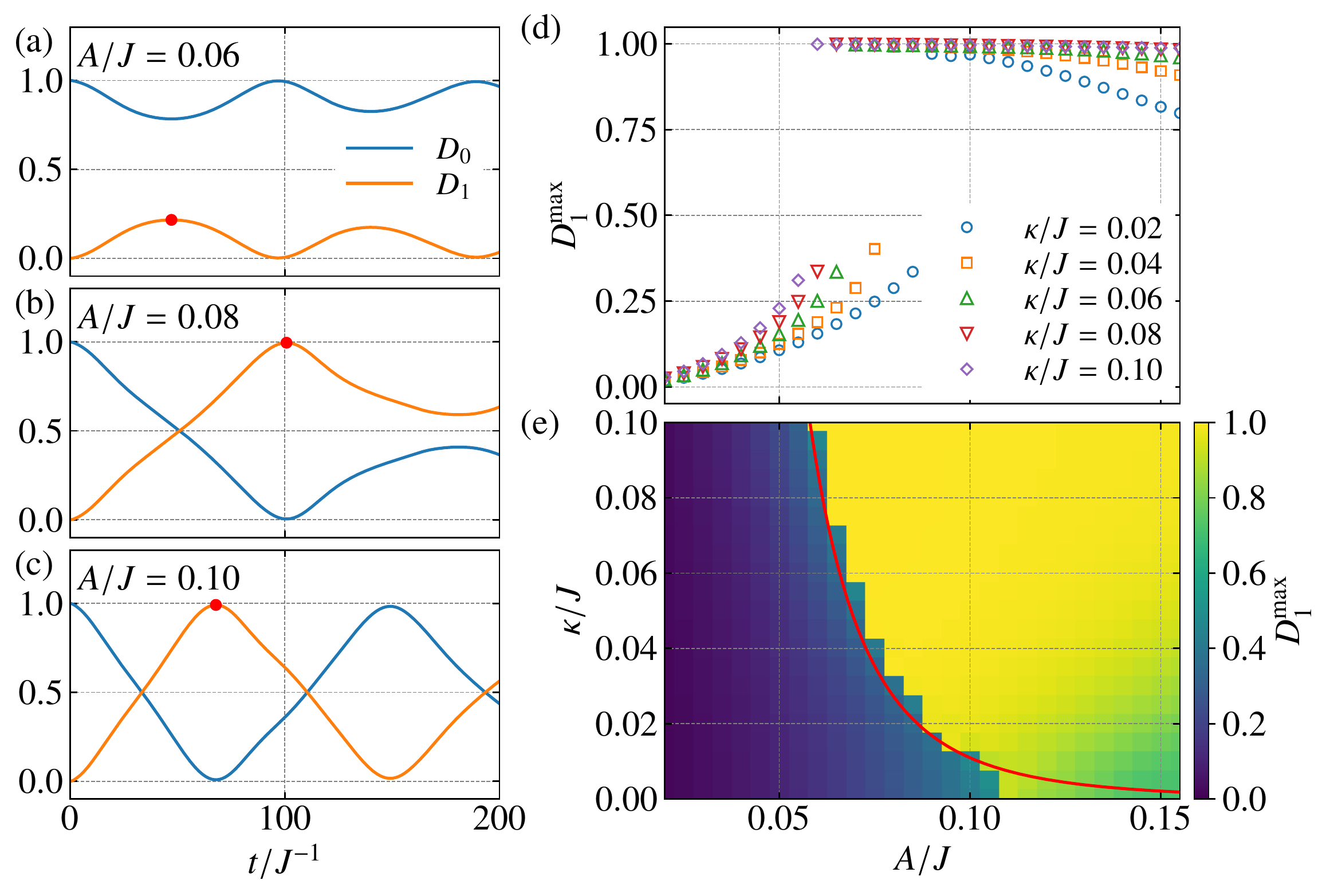}
  \caption{
    (a)--(c) Time evolution of the vison densities $D_p$ at $p=0,1$ after abruptly introducing the local field with (a) $A/J=0.06$, (b) $0.08$, and (c) $0.1$ at the site $j=1$ [see Fig.~\ref{fig:lattice}(b)].
    The filled circles denote the maximum of $D_1$.
    (d) $A$ dependence of the maximum $D_1^{\rm max}$ for several $\kappa$.
    (e) Contour plot of $D_1^{\rm max}$ on the $A$-$\kappa$ plane.
    The red line represents $A_{\rm cr}=0.032\kappa^{-0.25}$.
  }
  \label{fig:tdep}
  \end{center}
\end{figure}

To examine this feature in more detail, we introduce a simple setup, where a vison is located at $p=0$ in the initial state and the local magnetic field is applied to the spin at site $j=1$ as a quenched field with $h_1(t)=A\theta(-t)$ [see Fig.~\ref{fig:lattice}(b)].
Since $W_0$ and $W_1$ do not commute with $S_1^z$, their expectation values should change by the local field quench.
On the other hand, in the present initial condition, the relation $\means{W_0+W_1}=0$, namely $D_0+D_1=1$, is maintained even in the presence of the local field.
Figures~\ref{fig:tdep}(a)--\ref{fig:tdep}(c) show the time evolutions of $D_0$ and $D_1$ for $\kappa/J=0.05$ for several $A$.
In the case of $A/J=0.06$, $D_1$ takes a maximum $D_1^{\rm max}\simeq 0.2$ at $t/J^{-1}\simeq 50$ and turns to decrease and oscillate over time.
On the other hand, at $A/J=0.08$, the time evolution of $D_1$ exhibits a completely different behavior; the maximum value $D_1^{\rm max}$ reaches almost one for $A/J=0.08$, suggesting the hopping of a vison from $p=0$ to $p=1$ occurs.
Further increasing $A$, the time evolution of $D_0$ and $D_1$ appear to be cosine-type.
Note that $D_1$ takes a maximum at $t=t_{\rm max}\sim 69J^{-1}$ at $A/J=0.1$, corresponding to the time width $T$ of the local field for controlling a vison, used in Fig.~\ref{fig:XYmap}. 

\begin{figure}[t]
  \begin{center}
  \includegraphics[width=\columnwidth,clip]{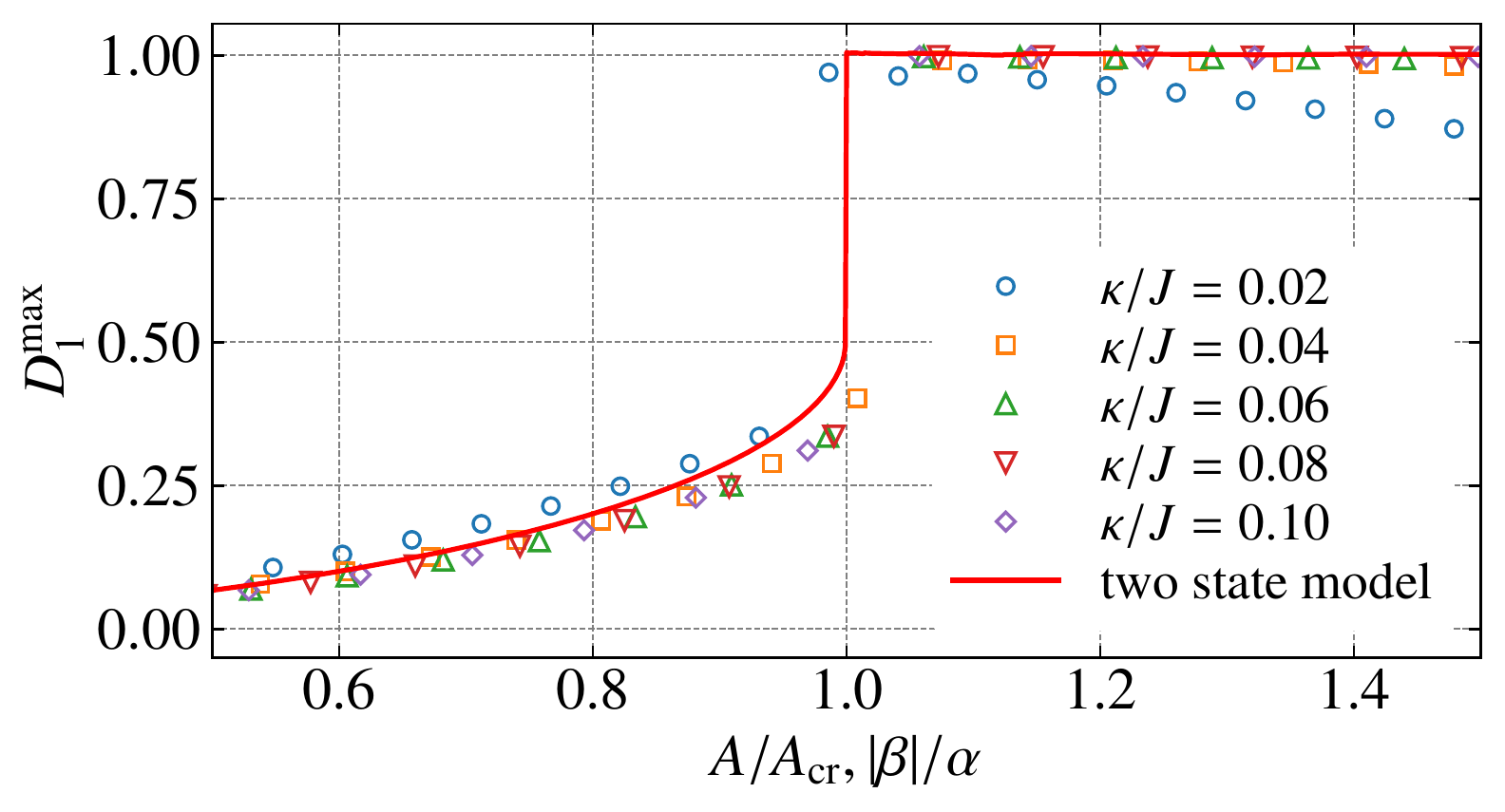}
  \caption{
    Plot of $D^{\rm max}$ scaled by $A_{\rm cr}$ for the horizontal axis in Fig.~\ref{fig:tdep}(d).
    $D^{\rm max}$ in the two-state model is also shown by a red line as a function of $|\beta|/\alpha$.
  }
  \label{fig:Adep}
  \end{center}
\end{figure}

We also find the existence of a threshold $A_{\rm cr}$ for the local field intensity; the vison hopping assisted by the local field does not occur below $A_{\rm cr}$ but the field is capable of manipulating visons above $A_{\rm cr}$.
The value of $A_{\rm cr}$ can be determined by observing the maximum of $D_1$, defined as $D_1^{\rm max}$.
Figures~\ref{fig:tdep}(d) and \ref{fig:tdep}(e) display $D_1^{\rm max}$ as a function of $A$ for several $\kappa$ and its contour map on the $A$-$\kappa$ plane~\footnote{We find that $D_1^{\rm max}$ decreases with increasing $A$ for small $\kappa$, as shown in Fig.~\ref{fig:tdep}(d).
Phenomenologically, this is considered to arise from Majorana fermions excited by strong magnetic fields because the behavior becomes less pronounced by an increase of $\kappa$ that produces a widening of the Majorana gap.}.
These results clearly show the presence of $A_{\rm cr}$ at which $D_1^{\rm max}$ abruptly changes to one.
Furthermore, $A_{\rm cr}$ decreases with increasing $\kappa$.
This result suggests that, if the background uniform field $\kappa$ is weak, a stronger local field $A$ is needed for the vison manipulation, but a weak local field can move the vison when the uniform field is strong.
This is a trade-off relation between the uniform field and local field intensities.
Moreover, we find $A_{\rm cr}\simeq 0.032\kappa^{-0.25}$ by fitting $A_{\rm cr}$, which is shown in Fig.~\ref{fig:tdep}(e) as a red line.
Using the values of $A_{\rm cr}$, we rescale the $A$ dependence of $D_1^{\rm max}$ as presented in Fig.~\ref{fig:Adep}.
All data almost collapse to a single curve, suggesting the presence of universal behavior in vison hoppings.


Here, we discuss the origin of the threshold behavior for the vison hopping triggered by the local field.
We construct a simplified low-energy model to consider the two states presented in Fig.~\ref{fig:lattice}(b), where the upper and lower states are denoted as $\kets{\Psi_0}$ and $\kets{\Psi_1}$, respectively.
The time-dependent wave function is written as $\kets{\Psi(t)}=c_0(t)\kets{\Psi_0} + c_1(t)\kets{\Psi_1}$ with the coefficients $c_0(t)$ and $c_1(t)$, whose time evolutions are determined by the Schr\"odinger equation $i\partial \kets{\Psi(t)}/\partial t= {\cal H}_{\rm low}\kets{\Psi(t)}$.
We assume the effective Hamiltonian ${\cal H}_{\rm low}$ as the following form:
\begin{align}
 {\cal H}_{\rm low}=
 \begin{pmatrix}
  -2\alpha\Delta(t) & \beta\\
 \beta^* & 2\alpha \Delta(t)
 \end{pmatrix},
\end{align}
where $\Delta(t)=|c_0(t)|^2-|c_1(t)|^2$, which causes a nonlinearity in the system.
The off-diagonal terms with the coefficient $\beta$ arise from the local magnetic field applied at site $j=1$, and $|\beta|$ is expected to be given by $A|\bras{\Psi_1}S_1^z\kets{\Psi_0}|$ with the local-field intensity $A$.
On the other hand, the diagonal term with the real coefficient $\alpha$ mainly originates from interactions between Majorana fermions and visons intrinsic in the Kitaev model, which causes vison localization ~\cite{Lahtinen2011}.
We impose $c_0(t=0)=1$ and $c_1(t=0)=0$ as initial conditions and calculate their time evolutions~\cite{suppl}.
Since $|c_p(t)|^2$ ($p=0,1$) is regarded as the probability of the vison existing at $p$, we assume $D_p=|c_p(t)|^2$, and $D_1^{\rm max}$ is determined similarly to the Hartree-Fock calculations.
This quantity is plotted as a function of $|\beta|/\alpha$ in Fig.~\ref{fig:Adep}(b) as a red line, which exhibits a jump at $|\beta|/\alpha=1$.
The line well coincides with the Hartree-Fock results scaled by $A_{\rm cr}$, implying that the low-energy two-state model can well account for the vison hopping in the lattice system.

In the case with $|\beta|/\alpha\gg1$, the diagonal term giving the nonlinearity can be neglected and $|c_1(t)|^2=\sin^2 |\beta| t$, indicating $t_{\rm max}=\pi/(2|\beta|)$.
The previous studies suggested $|\beta|/A=|\bras{\Psi_1}S_1^z\kets{\Psi_0}|\simeq 0.23$ at $\kappa/J=0.05$~\cite{Joy2022,Chen2023}, leading to $t_{\rm max}/J^{-1}\simeq 68$ for $A/J=0.1$.
The estimation of $t_{\rm max}$ is consistent with the Hartree-Fock result shown in Figs.~\ref{fig:XYmap} and \ref{fig:tdep}(c).
Thus far, we assume that the local magnetic field is constant for $t$.
Even if one considers the time-dependent local field $h_1(t)$, which gives rise to the $t$ dependence for $\beta$, the condition for $|c_1(t_{\rm max})|^2=1$ is determined only by the integral as $\int_{t_{\rm in}}^{t_{\rm max}} |\beta(t)| dt=\pi/2$.
The result will offer a simple guideline for manipulating a vison.

\begin{figure}[t]
  \begin{center}
  \includegraphics[width=\columnwidth,clip]{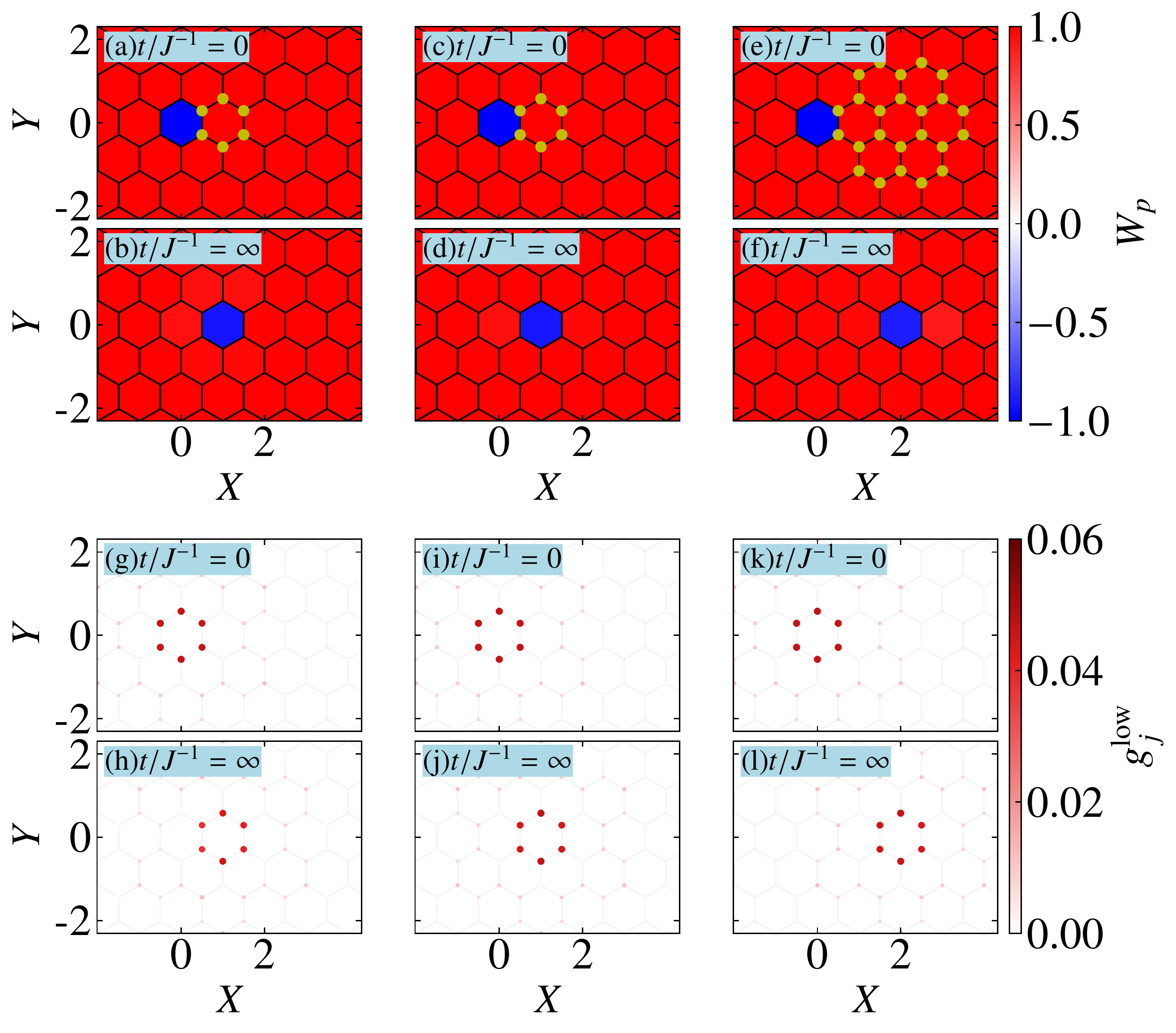}
  \caption{
    (a),(b) Spatial maps of $W_p$ (a) before and (b) after introducing the time-dependent magnetic field $h_j(t)$ applied for the six yellow sites shown in (a) as a rectangular function $A \mathcal{R}(t;0,T)$ with $A/J=0.04$ and $T/J^{-1}=171$.
    (c)--(f) Same plots for $h_j(t)$ applied to (c),(d) the six and (e),(f) the twenty-four yellow sites as a Gaussian function $A e^{-(t-t_c)^2/2\sigma^2}$ with $A/J=0.08$, $\sigma/J=100$ and $t_c/J^{-1}=300$.
    (g)--(l) Corresponding plots to (a)--(f) for low-energy LDOS $g_j^{\rm low}$.
    In these simulations, we set $\kappa/J=0.05$.
  }
  \label{fig:variations}
  \end{center}
\end{figure}

To advance the practical application of vison manipulation, we investigate several variations of local fields.
As illustrated in Fig.~\ref{fig:variations}, a localized vison can be moved by the local fields applied to neighboring six sites as rectangular and Gaussian functions and also by the field applied even to adjacent twenty-four sites. 
These calculations suggest that visons are not additionally excited by the local fields with spatial distributions.
This is interpreted as follows.
When applying a local magnetic field to a specific site adjacent to a vison, the vison moves to the neighboring plaquette because the local magnetic field hybridizes two different states whose energies are degenerate. 
In contrast, a local field applied at a site without visons possibly creates a pair of two visons in addition to an existing one, which demands a nonzero excitation energy~\cite{Kitaev2006}.
Therefore, even if local magnetic fields with spatial distributions are employed beyond the field at a specific site, the excitation gap prevents the creation of visons and ensures the manipulation of a vison while preserving its shape.
Furthermore, we have verified that a Majorana zero mode is always accompanied by the vison motions, as depicted in Figs.~\ref{fig:variations}(g)--\ref{fig:variations}(l).
These results imply that the Majorana zero mode can be manipulated even by using local magnetic fields applied to multiple sites in wider regions beyond the application only to a specific site.

Within our framework, the required local-field strength and the upper limit of its sweeping speed are estimated at $\sim 1~{\rm T}$ and $\sim 10~{\rm m/s}$, respectively, for the Kitaev candidate material $\alpha$-RuCl$_3$ with $J\simeq 100$~K ~\cite{Banerjee2016_NatMate,Winter2016,Do2017majorana,hirobe2017} and $\kappa/J\simeq 0.05$ ~\cite{Tanaka2022}.
While the field strength is larger than that given by recent STM technique generating nano-scale magnetic fields at present ~\cite{Singha2021localmag}, our study will provide a possible route to spatiotemporal manipulation of a vison and stimulate the development of experimental research.

Here, we propose several routes to reduce the required magnetic field.
Based on the present study, we find that $A_{\rm cr}$ decreases with increasing $\kappa$ as shown in Fig.~\ref{fig:tdep}(e), suggesting that the increase of the uniform effective field results in the reduction of the required field strength.
We have also confirmed that extending the region of an applied local field also enables manipulating a vison in lower local fields.
This phenomenon originates from the presence of multiple sites at which the local fields contribute to the hopping of a vison~\cite{Joy2022,Chen2023}.
Furthermore, selecting candidate materials with smaller energy scales of the Kitaev interaction~\cite{Yamada2017,Jang2020A2PrO3} is also effective to reduce the the required field strength.



In summary, we have demonstrated that a vison accompanied by a Majorana zero mode can be manipulated spatiotemporally by a locally applied magnetic field sweeping in the Kitaev model under a uniform effective field.
Increasing the intensities of the local and uniform fields enhances controllability due to interactions between visons and itinerant Majorana fermions.
Notably, the vison gap plays a crucial role in the stable manipulation of a vison. 
Our finding sheds light on the impact of many-body effects of fractional quasiparticles on the vison manipulation and paves the way for achieving topological quantum computation in Kitaev spin liquids.

\begin{acknowledgments}
Parts of the numerical calculations were performed in the supercomputing systems in ISSP, the University of Tokyo.
This work was supported by Grant-in-Aid for Scientific Research from
JSPS, KAKENHI Grant Nos.~JP19K03742, JP20H00122, JP20K14394, JP23K13052, and JP23H01108.
It is also supported by JST PRESTO Grant No.~JPMJPR19L5. 
\end{acknowledgments}

\bibliography{refs}

\clearpage
\onecolumngrid

\appendix
\vspace{15pt}
\begin{center}
{\large \bf Supplemental Material for ``Field-driven spatiotemporal manipulation of Majorana zero modes in Kitaev spin liquid''}
\end{center}

\setcounter{figure}{0}
\setcounter{equation}{0}
\setcounter{table}{0}
\renewcommand{\thefigure}{S\arabic{figure}}
\renewcommand{\theequation}{S\arabic{equation}}
\renewcommand{\thetable}{S\Roman{table}}
\baselineskip=6mm

\section{Details of time-dependent Hartree-Fock theory}

In this section, we introduce the time-dependent Hartree-Fock theory used in the present calculations.
In the absence of the local field, the Kitaev model described by ${\cal H}_{K}$ can be solved exactly as it is given by a bilinear form of $c_j$ under a fixed configuration of the local conserved quantities $\eta_r$ consisting of $\bar{c}_j$.
On the other hand, the external field ${\cal H}_h(t)$ mixes the two Majorana fermions $c_j$ and $\bar{c}_j$, which gives rise to the dynamics of $\bar{c}_j$.
Namely, $\eta_r$ are no longer local conserved quantities.
In this case, one needs to treat the third, fourth, and fifth terms in Eq.~(2) in the main text as interactions between two Majorana fermions $c_j$ and $\bar{c}_j$.
Here, we adopt the mean-field theory for the Majorana fermion representation in ${\cal H}={\cal H}_K+{\cal H}_h(t)$ by applying Hartree-Fock decouplings to these terms as 
\begin{align}
i\bar{c}_{i}\bar{c}_{j}ic_{i}c_{j}\simeq
\means{i\bar{c}_{i}\bar{c}_{j}}ic_{i}c_{j}+i\bar{c}_{i}\bar{c}_{j}\means{ic_{i}c_{j}} -\means{ic_{i}\bar{c}_{j}}ic_{j}\bar{c}_{j}
-ic_{i}\bar{c}_{j}\means{ic_{j}\bar{c}_{j}}+\means{ic_{i}\bar{c}_{j}}ic_{j}\bar{c}_{i}+ic_{i}\bar{c}_{j}\means{ic_{j}\bar{c}_{i}} + {\rm const.}
\end{align}
The obtained mean-field Hamiltonian ${\cal H}_{\rm MF}(t)$ is given by a following bilinear form of Majorana fermions $\{\gamma_l\}=\{c_{1},c_{2},\cdots,c_{N},\bar{c}_{1},\bar{c}_{2},\cdots,\bar{c}_{N} \}$:
\begin{align}
  \mathcal{H}_{\rm MF} = \frac{i}{4}\sum_{ll'}\gamma_{l}{\cal A}_{ll'}\gamma_{l'}+C,
\end{align}
where $N$ is the number of sites, $C$ is a constant, and ${\cal A}$ is a $2N\times2N$ real skew symmetric matrix.
Note that ${\cal A}$ is a function of $t$ and mean fields $\means{\gamma_l \gamma_{l'}}$, where the expectation values are taken for many-body wave function $\kets{\Psi(t)}$.
This wave function is determined by the Schr\"odinger equation as follows:
\begin{align}
  i\frac{\partial}{\partial t} \kets{\Psi(t)}={\cal H}_{\rm MF}(t)\kets{\Psi(t)},\label{eq:S-schrodinger}
\end{align}
where $\hbar$ is assumed to be unity.
Here, we define the density matrix whose component given by
\begin{align}
  \rho_{ll'}=\frac{1}{4}(\bra{\Psi(t)}\gamma_{l'}\gamma_{l}\ket{\Psi(t)}-\delta_{ll'}).
\end{align}
The density matrix $\rho$ is a $2N\times2N$ Hermitian matrix, which is regarded as a set of mean fields.
From Eq.~\eqref{eq:S-schrodinger}, one can find that the density matrix $\rho$ obeys the following von Neumann equation:
\begin{align}
  i\frac{\partial \rho}{\partial t} =[i{\cal A},\rho].
\end{align}
We calculate the time evolution based on this equation.
Since ${\cal A}$ is a function of $t$ and $\rho$, we use the fourth-order Runge-Kutta method with the step size $\Delta t$ to compute the time evolution.

As an initial state at $t_{\rm in}=0$, we impose no local magnetic field with ${\cal H}_h(t_{\rm in})=0$.
At $t=t_{\rm in}$, $W_p$ are conserved quantities, and hence, we can assume any configurations of vison excitations as an initial state $\kets{\Psi(t_{\rm in})}$ and easily obtain the matrix elements of $\rho(t=t_{\rm in})$ by diagonalizing $i{\cal A}$ at $t=t_{\rm in}$.
Since ${\cal A}$ is a real skew-symmetric matrix, the eigenvalues of $i{\cal A}$ appear as pairs of real numbers with opposite signs.
The positive eigenvalues are written as $\varepsilon_\lambda$ with $\lambda=1,2,\cdots,N$.
The diagonalization of $i{\cal A}$ is performed by the $2N\times 2N$ matrix $U$ as
\begin{align}
  U^\dagger i{\cal A}U={\rm diag}\{\varepsilon_1,\varepsilon_2,\cdots, \varepsilon_N,-\varepsilon_1,-\varepsilon_2,\cdots, -\varepsilon_N\}.
\end{align}
Here, we introduce $2N\times N$ matrix ${\cal U}$, which is defined as
\begin{align}
  U=\begin{pmatrix}
    {\cal U} & {\cal U}^*
  \end{pmatrix},
\end{align}
where the relations ${\cal U}^\dagger {\cal U}=1$ and ${\cal U}^T {\cal U}=0$ are imposed.
To satisfy these conditions even in the presence of a zero-energy degeneracy, we perform the numerical diagonalization using the Schur decomposition for ${\cal A}$.
We introduce fermions $f_\lambda$ as
\begin{align}
  c_j=\sqrt{2}\sum_{\lambda}\left({\cal U}_{j\lambda}f_\lambda+{\cal U}_{j\lambda}^* f_\lambda^\dagger\right).\label{eq:Scj}
\end{align}
Then, the mean-field Hamiltonian at $t=t_{\rm in}$ is written as
\begin{align}
  \mathcal{H}_{\rm MF}(t=t_{\rm in}) = \sum_{\lambda}\varepsilon_\lambda \left(f_\lambda^\dagger f_\lambda+\frac{1}{2}\right)+C.
\end{align}

Next, we introduce the local density of states (DOS) $g_j^0(\varepsilon)$ for the itinerant Majorana fermion $c_j$ at site $j$ at the initial time $t=t_{\rm in}$ as
\begin{align}
  g_j^0(\varepsilon)=\sum_\lambda |\bras{0}c_j f_\lambda^\dagger \kets{0}|^2\delta(\varepsilon-\varepsilon_\lambda),
\end{align}
where $\kets{0}$ is the vacuum satisfying $f_\lambda\kets{0}=0$. 
From Eq.~\eqref{eq:Scj}, this is rewritten as
\begin{align}
  g_j^0(\varepsilon)=2\sum_\lambda |{\cal U}_{j\lambda}|^2\delta(\varepsilon-\varepsilon_\lambda).
\end{align}
The local DOS at $t>t_{\rm in}$ is introduced by a similar manner as
\begin{align}
  g_j(\varepsilon;t)=2\sum_\lambda |{\cal U}_{t;j\lambda}|^2\delta(\varepsilon-\varepsilon_{t;\lambda}),
\end{align}
where $\varepsilon_{t;\lambda}$ is the eigenvalue for $\mathcal{H}_{\rm MF}(t)$ with the transformation matrix ${\cal U}_t$.
Note that the following sum rule is satisfied:
\begin{align}
  \int_0^\infty g_j(\varepsilon;t) d\varepsilon =1.
\end{align}

\section{Honeycomb lattice cluster}

\begin{figure}[t]
  \begin{center}
  \includegraphics[width=\columnwidth,clip]{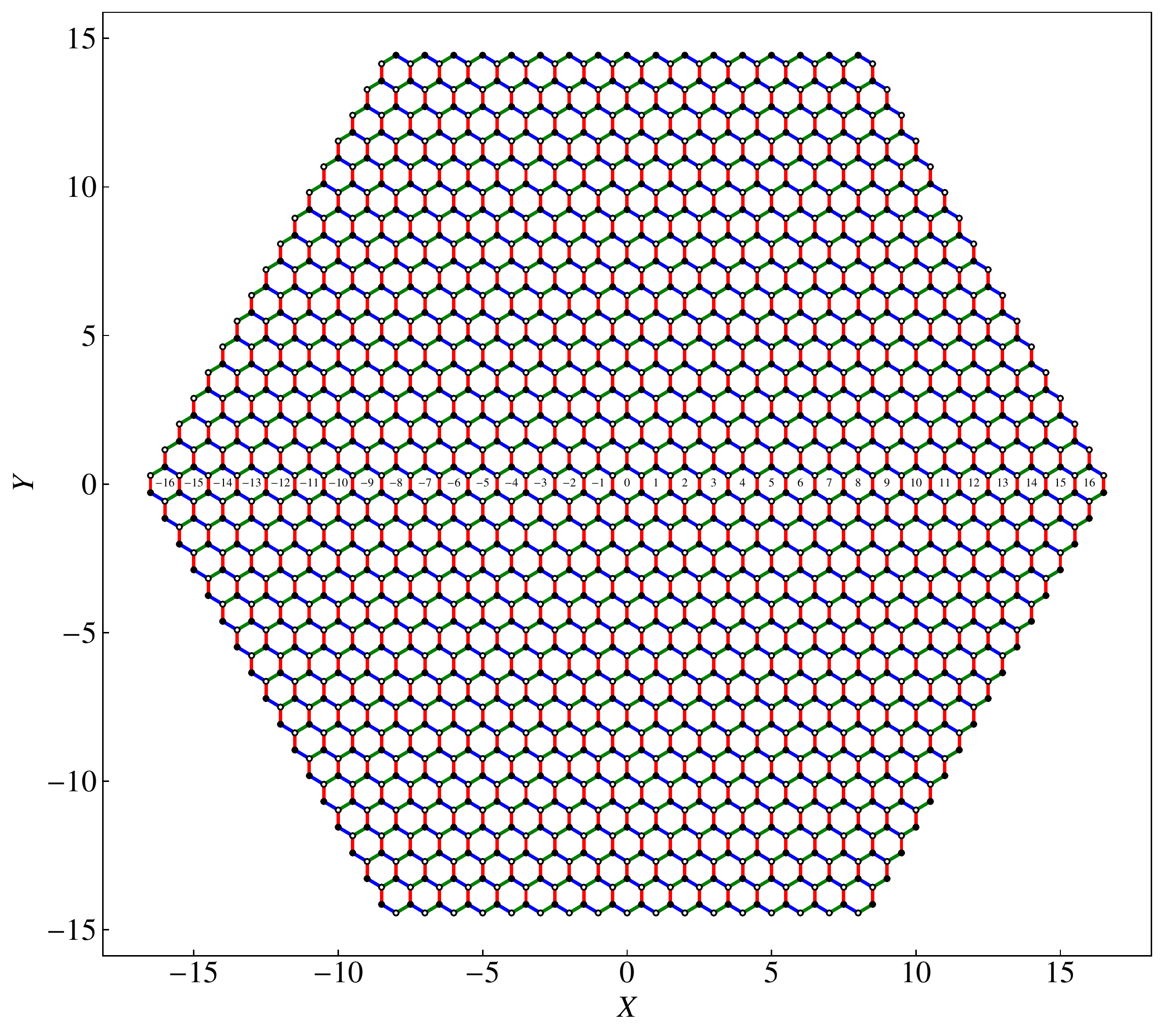}
  \caption{
Honeycomb lattice cluster with $N=1734$ used in the calculations presented in the main text.
The numbering for the hexagon plaquettes with centers on the $Y=0$ line is displayed.
  }
  \label{figS:cluster}
  \end{center}
\end{figure}

Figure~\ref{figS:cluster} shows the hexagon-type cluster with $N=1734$ used in the present calculations.
This is straightforwardly obtained by extending the cluster shown in Fig.~1(a).

\section{Local-field dependence of vison transport}

\begin{figure}[t]
  \begin{center}
  \includegraphics[width=\columnwidth,clip]{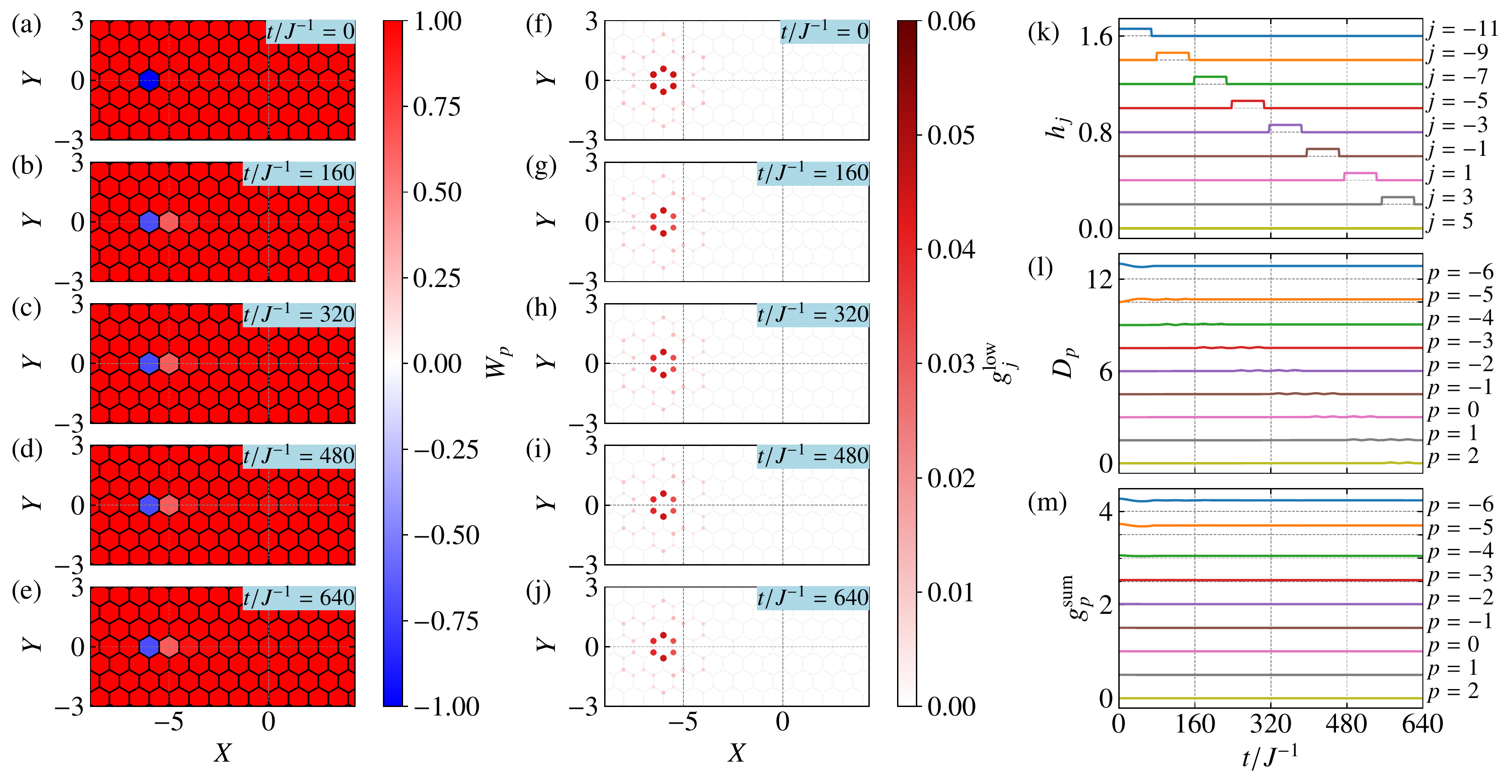}
  \caption{
    Corresponding plots to Fig.~2 in the main text for the case with $A/J=0.06$.
  }
  \label{figS:XYmap0.06}
  \end{center}
\end{figure}

\begin{figure}[t]
  \begin{center}
  \includegraphics[width=\columnwidth,clip]{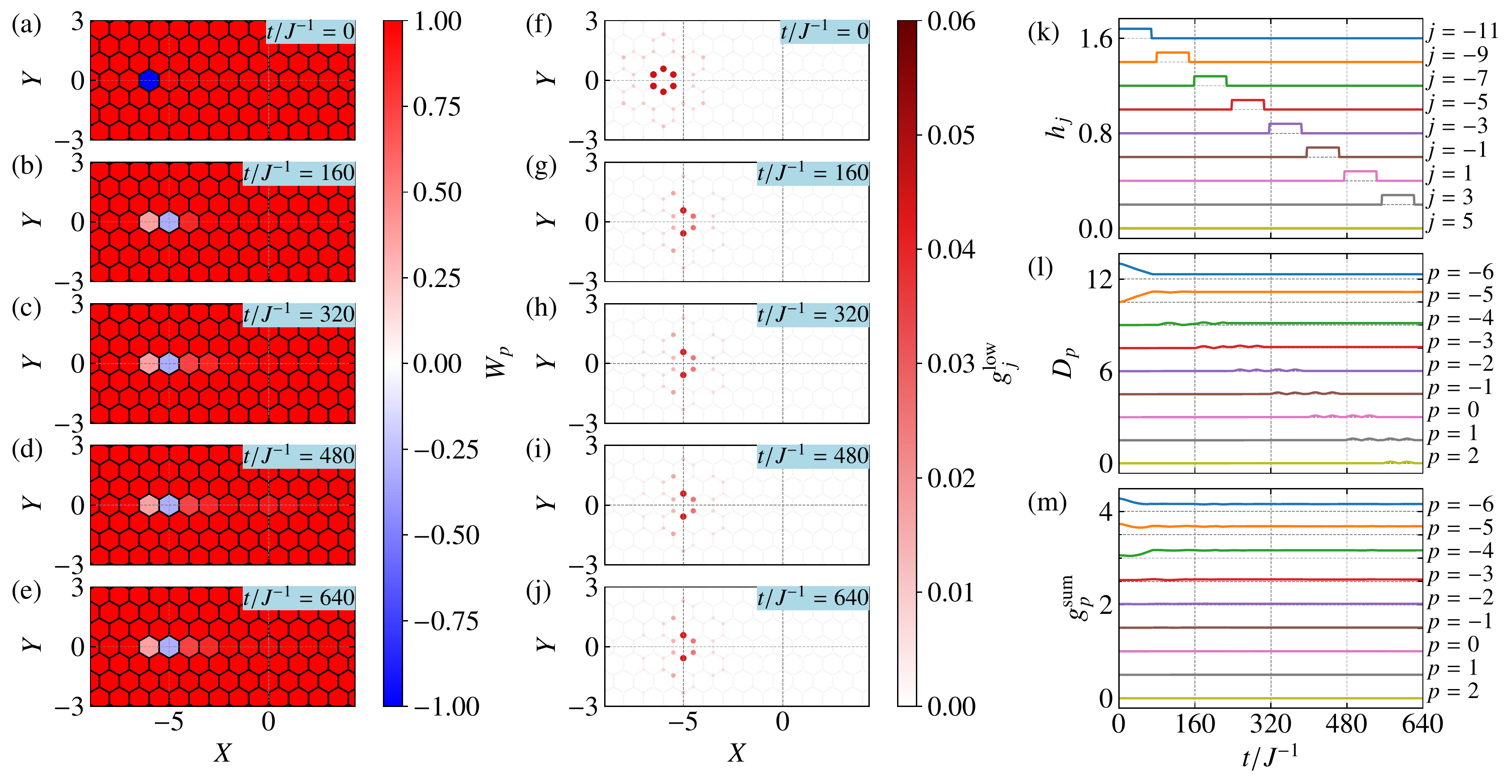}
  \caption{
    Corresponding plots to Fig.~2 in the main text for the case with $A/J=0.08$.
  }
  \label{figS:XYmap0.08}
  \end{center}
\end{figure}

In this section, we present results by varying local field intensities $A$.
Figures~\ref{figS:XYmap0.06} and ~\ref{figS:XYmap0.08} show corresponding plots to Fig.~2 in the main text for the case with $A/J=0.06$ and $A/J=0.08$, respectively.
In these cases, a vison does not follow the motion of the local field. 

\section{Time evolution in low-energy effective model}

In this section, we present results for time evolution in a low-energy two-state model with $\kets{\Psi_0}$ and $\kets{\Psi_1}$, which is introduced in the main text.
The wave function is given as a superposition of these states: $\kets{\Psi(t)}=c_0(t)\kets{\Psi_0} + c_1(t)\kets{\Psi_1}$.
We assume the Hamiltonian to be the following matrix form for the two states:
\begin{align}
  {\cal H}_{\rm low}=
  \begin{pmatrix}
   -2\alpha\Delta(t) & \beta\\
  \beta^* & 2\alpha \Delta(t)
  \end{pmatrix},
 \end{align}
 where $\Delta(t)=|c_0(t)|^2-|c_1(t)|^2$, and $\alpha$ and $\beta$ are real and complex constants, respectively. 
We calculate the time evolution based on the Schr\"odinger equation,
\begin{align}
  i\frac{\partial}{\partial t}
  \begin{pmatrix}
    c_0(t)\\c_1(t)
  \end{pmatrix}
  =
  {\cal H}_{\rm low}
  \begin{pmatrix}
    c_0(t)\\c_1(t)
  \end{pmatrix},
\end{align}
under the initial conditions with $c_0=1$ and $c_1=0$ at $t_{\rm in}=0$.

\begin{figure}[t]
  \begin{center}
  \includegraphics[width=0.5\columnwidth,clip]{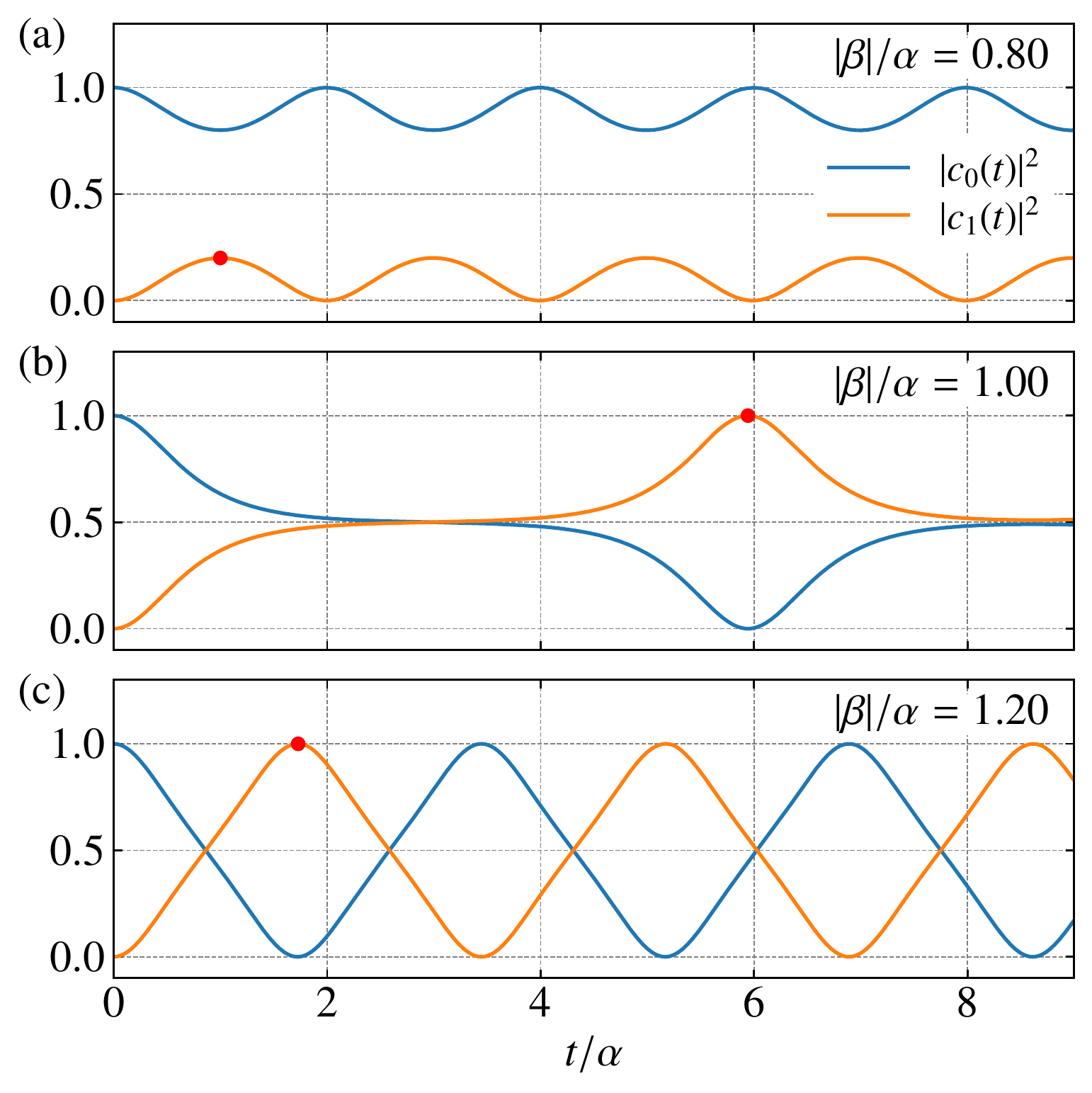}
  \caption{
Time evolutions of $|c_0(t)|^2$ and $|c_1(t)|^2$ in the two-state model with (a) $|\beta|/\alpha=0.8$, (b) $|\beta|/\alpha=1$, and (c) $|\beta|/\alpha=1.2$.
The filled circles denote the first maximum of $|c_1(t)|^2$.
  }
  \label{figS:tsm}
  \end{center}
\end{figure}

Figure~\ref{figS:tsm} presents the time evolutions of $|c_0(t)|$ and $|c_1(t)|$ for several values of $|\beta|/\alpha$.
Below $|\beta|/\alpha=1$, $|c_0(t)|^2$ and $|c_1(t)|^2$ oscillate around one and zero, respectively, and $|c_1(t)|^2$ does not become one [Fig.~\ref{figS:tsm}(a)].
One the other hand, above $|\beta|/\alpha=1$, $|c_0(t)|^2$ and $|c_1(t)|^2$ oscillate between zero and one, as shown in Fig.~\ref{figS:tsm}(c).
At $|\beta|/\alpha=1$ corresponding to the critical point, $|c_1(t)|^2$ reaches one although the period is much longer than that of the other parameters.

\end{document}